# Higher order transmission resonances in above-barrier reflection of ultra-cold atoms

H.A. Ishkhanyan[1], V.P. Krainov[1], and A.M. Ishkhanyan[2]

[1]*Moscow Institute of Physics and Technology, 141700 Dolgoprudny, Russia*
[2]*Institute for Physical Research NAS of Armenia, 0203 Ashtarak-2, Armenia*

**Abstract.** The reflectionless transmission resonances in above-barrier reflection of Bose-Einstein condensates by the Rosen-Morse potential are considered using the mean field Gross-Pitaevskii approach. Applying an exact third order nonlinear differential equation obeyed by the condensate's density, the exact solution of the problem for the first resonance is derived. It is shown that in the nonlinear case the total transmission is possible for positive potential heights, i.e., for potential barriers. Further, it is shown that an appropriate approximation for higher-order resonances can be constructed using a limit solution of the equation for the density written as a root of a polynomial equation of the third degree. Using this limit function and the solution for the first resonance, a simple approximation for the shift of the nonlinear resonance potential's depth from the corresponding linear resonance's position is constructed for higher order resonances. The result is written as a linear function of the resonance order. This behavior notably differs from the case of the rectangular barrier for which the nonlinear shift is approximately constant for all the resonance orders.

## Introduction

The basic effects of tunneling through a barrier and above-barrier reflection of particles are well appreciated paradigms of quantum mechanics extensively studied for many decades [1]. The experimental realization of Bose-Einstein condensation in dilute gases of ultracold atoms [2] has stimulated a renewed interest to those effects since the condensates provide a different test of fundamental principles of quantum mechanics due to the essentially nonlinear nature of the many-body dynamics of Bose-condensates [3]. Several physical situations for which the linear analogue is known have recently been discussed including, e.g., the step-potential [4], the rectangular barrier [5], the double-delta-shell and delta-comb configurations [6]. The above-barrier reflection of cold atoms by the squared-sech Rosen-Morse potential has been addressed in [7,8,9]. The exact solution of the Gross-Pitaevskii equation has been reported for a certain depth of the potential [8] and the reflection coefficient has been calculated for the close vicinity of the first linear transmission resonance for the case of small nonlinearity [9]. In the present paper, we consider the higher order transmission resonances viewed in terms of incoming and outgoing waves.

The dynamics of Bose-Einstein condensates in the mean-field approximation is described by a nonlinear Schrödinger equation referred to as the Gross-Pitaevskii equation [3,10]. In the one-dimensional case this equation is written as

$$i\hbar \frac{\partial \Psi}{\partial t} = -\frac{\hbar^2}{2m}\frac{\partial^2 \Psi}{\partial x^2} + (V(x) + g|\Psi|^2)\Psi = 0, \qquad (1)$$

where the nonlinearity parameter $g$ determines the mean-field self-interaction and $V(x)$ is the reflecting potential. Applying the ansatz $\Psi(x,t) = \exp(-i\mu t/\hbar)\psi(x)$, where $\mu$ is the chemical potential, Eq. (1) is reduced to the following stationary version

$$-\frac{1}{2}\frac{d^2\psi}{dx^2} + (-\mu + V(x) + g|\psi|^2)\psi = 0. \qquad (2)$$

The reflecting potential we consider here is the finite-height squared hyperbolic-secant Rosen-Morse potential

$$V(x) = V_0 \operatorname{sech}^2 x. \qquad (3)$$

To meet the conditions for *above*-barrier reflection, we suppose that $\mu > 0$ for a potential well with $V_0 < 0$ and $\mu > V_0$ for a barrier with $V_0 > 0$. Further, to treat the process in terms of incoming and outgoing waves, we look for a solution of Eq. (2) which at $x \to -\infty$ has the asymptotic behavior of the form $\psi \sim C_1 e^{+ikx} + C_2 e^{-ikx}$ describing a combination of incident and reflected running waves, and at $x \to +\infty$ has the asymptote $\psi \sim C_3 e^{+ikx}$ corresponding to a transmitted wave. The reflectionless transmission then corresponds to a specific case when the reflected wave is absent, i.e., $C_2 = 0$ and, hence, $|C_1| = |C_3|$. Substituting the ansatz $\psi = C_1 e^{+ikx}$ into Eq. (2) and noting that the Rosen-Morse potential vanishes at $x \to -\infty$, we readily see that the wave number $k$ is determined as $k = \sqrt{2(\mu - |C_1|^2 g)}$. Hence, since $k$ is supposed to be a real number, it should be $\mu > |C_1|^2 g$. Note finally that the normalization of the wave function can always be incorporated in the definition of the nonlinearity parameter $g$. For this reason, without loss of generality, we put $|C_1| = 1$ and thus adopt the normalization $|\psi(-\infty)|^2 = 1$.

In the linear case $g = 0$ the reflectionless transmission is known to occur for a discrete spectrum of *negative* values of the potential height defined as [1]

$$V_{Ln} = -\frac{1}{2}n(n+1), \quad n = 1, 2, 3, \ldots \qquad (4)$$

The corresponding solution of the linear Schrödinger equation for these transmission resonances satisfying the adopted normalization is written in terms of the Gauss hypergeometric function [11]:

$$\psi_{Ln} = e^{ikx} \,_2F_1(-n, 1+n, 1+ik_0, z), \quad k_0 = \sqrt{2\mu}, \tag{5}$$

where $z = (1+\text{th}\,x)/2$. Below we consider the transmission resonances for the nonlinear case $g \neq 0$. We adopt the numeration of the successive resonances for which the height of the $n$th order nonlinear resonance $V_{NLn}$ at $g \to 0$ tends to the height of the $n$th order linear resonance $V_{Ln}$ given by Eq. (4). Note that, in accordance with this formula, the reflectionless transmission in the linear case is possible only for potential wells. However, we will see below that in the nonlinear case the total transmission is also possible for positive potential heights, i.e., for potential barriers. Note also that the resonance position in the linear case does not depend on the chemical potential $\mu$. We will see that in the nonlinear case the situation is changed: the spectrum of the nonlinear resonances becomes a function of $\mu$.

**Equation for the probability and the limit solution**

An initial observation serving as a starting point for what follows is that the direct linearization of the stationary Gross-Pitaevskii equation (2) using the linear solution (5) as the zero-order approximation (i.e., effectively, a linearization via replacement of the $|\psi|^2$ term in Eq. (2) by $|\psi_{Ln}|^2$) does not produce accurate results. More advanced approaches include the multiple scale analysis [12] (we have applied this method to treat the first nonlinear resonance for the rectangular barrier [5] and the Rosen-Morse potential [9]) and the renormalization technique (this reveals that one should apply the linear solution using $k = \sqrt{2(\mu-g)}$ instead of $k_0 = \sqrt{2\mu}$ in Eq. (5)). Here, we suggest an alternative approach based on an exact third order differential equation for the probability $p = |\psi(x)|^2$. This equation is derived by substituting $\psi(x) = \psi_1(x) + i\psi_2(x)$ into Eq. (2). Separating then the real and imaginary parts, we get

$$\begin{aligned} -\frac{\psi_1''}{2} + [-\mu + V + g(\psi_1^2 + \psi_2^2)]\psi_1 &= 0, \\ -\frac{\psi_2''}{2} + [-\mu + V + g(\psi_1^2 + \psi_2^2)]\psi_2 &= 0, \end{aligned} \tag{6}$$

where the prime denotes differentiation with respect to $x$. Multiplying now the first equation by $\psi_1'$ and the second equation by $\psi_2'$ and summing the two equations leads to the equation

$$-\frac{1}{2}\frac{d}{dx}[(\psi_1')^2+(\psi_2')^2]+(-\mu+V+g\,p)p'=0. \tag{7}$$

Another equation is derived by multiplying the first equation of system (6) by $\psi_1$ and the second one by $\psi_2$, summing the equations and applying the identity $\psi_1\psi_1''+\psi_2\psi_2''=-(\psi_1'^2+\psi_2'^2)+p''/2$. The result reads

$$-\frac{p''}{4}+\frac{1}{2}(\psi_1'^2+\psi_2'^2)+(-\mu+V+g\,p)p=0. \tag{8}$$

The last step is straightforward. We differentiate Eq. (8) and use Eq. (7) to arrive at the third order differential equation

$$\frac{d}{dx}\left[-\frac{p''}{4}+(-\mu+V+g\,p)p\right]+(-\mu+V+g\,p)p'=0. \tag{9}$$

Since we are studying the reflectionless transmission through the barrier the boundary conditions we here impose are

$$p(-\infty)=p(+\infty)=1. \tag{10}$$

Note that such boundary conditions define a nonlinear eigenvalue problem for the derived equation (9). Finally, note that, since we look for a solution with traveling-wave-like asymptotic behavior $\psi(-\infty) \sim e^{ikx}$, the function $p(x)$ should additionally obey the conditions

$$p'(-\infty)=0, \quad p''(-\infty)=0. \tag{11}$$

The derived equation (9) turns out to be a productive one since it allows one to linearize the problem using the solution of the linear problem as a starting approximation. A simpler alternative possibility (dealing with elementary functions) suggested by this equation for a relatively large chemical potential $\mu$ is to treat the problem starting from a limit solution that is obtained when neglecting the second derivative term in square brackets in Eq. (9). Numerical simulations show that this approach is valid as far as $\mu > 0.25$. The development of this approach is useful since it is not based on the solution of the associated linear problem which is known only for a few cases. Instead, it works with a simple approximate solution that is readily written for any potential.

Generally speaking, dropping out the highest derivative term of a differential equation is a singular procedure because the reduced equation becomes of a lower order and, hence, the solution to this equation in general can not satisfy all the initial conditions (see, e.g., numerous examples discussed in Ref. [12]). However, in our particular case we obtain a solution that incidentally satisfies all the imposed boundary conditions. This is not a unique case. A similar favorable situation is faced, e.g., in treating the Landau-Zener transition in cold molecule formation via photoassociation of an atomic Bose-Einstein condensate [13].

According to aforesaid, the limit solution $p_0(x)$ is a function defined as the solution of the reduced equation

$$\frac{d}{dx}[(-\mu+V+gp)p]+(-\mu+V+gp)p'=0. \qquad (12)$$

This equation is straightforwardly integrated producing a cubic polynomial equation for $p_0$:

$$C_0+(-\mu+V+gp_0)p_0^2=0, \qquad (13)$$

where $C_0$ is the integration constant. The boundary condition $p(-\infty)=1$ gives $C_0=\mu-g$. The limit function defined by this equation is a symmetric [ $p_0(x)=p_0(-x)$ ], non-oscillatory function of $x$. The comparison of this function with the exact numerical result for the probability $p$ is shown in Fig. 1 under both resonant (Fig. 1a) and non-resonant (Fig. 1b) conditions (the incoming wave is supposed to move from right to left, the large-amplitude oscillations seen for large positive $x$ in Fig. 1b describe the interference of incident and reflected waves). It is seen that in both cases the limit function satisfactory approximates the behavior of the system in a space region embracing the potential, while it becomes invalid for $x \to +\infty$ if the potential depth is not resonant. Hence, we conclude that this function may be specifically useful for treatment of the reflectionless transmission resonances. We will show below that this is indeed the case.

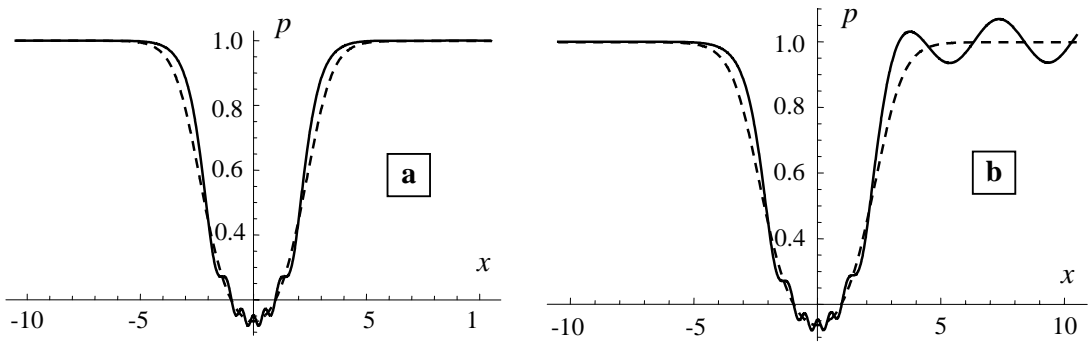

Fig. 1. Comparison of the limit function $p_0(x)$ defined by Eq. (13) (dotted line) with the exact numerical result for $p$ (solid line) for $\mu=0.5$ and $g/\mu=0.25$ (the incoming wave is supposed to move from right to left). a) The depth of the potential is chosen to support the sixth-order transmission resonance: $V_0=V_{NL6}=-20.4914$. b) The depth of the potential here is $V_0=-20$; it does not meet the transmission resonance conditions. The large-amplitude oscillations seen for large positive $x$ describe the interference of incident and reflected waves.

**Exact solution for the first nonlinear transmission resonance**

The equation (9) for probability $p$ allows deriving the exact solution describing the first nonlinear transmission resonance. This solution is obtained when searching of a particular solution of the form

$$p = 1 + a\,\text{sech}^2 x. \qquad (14)$$

To simplify the calculations, we rewrite Eq. (9) in the form

$$\frac{d}{dx}\left[-\frac{p''}{4} + \left(-2\mu p + Vp + \frac{3}{2}g p^2\right)\right] + V p' = 0 \qquad (15)$$

and recall that the second derivative of function $f = \text{sech}^2 x$ is written as a quadratic polynomial in $f$:

$$f'' = 4 f\left(1 - \frac{3}{2}f\right). \qquad (16)$$

Eq. (15) is then written as

$$\frac{d}{dx}\left[-af\left(1 - \frac{3}{2}f\right) + \left((-2\mu + V_0 f)(1 + a f) + \frac{3}{2}g(1 + a f)^2\right) + aV_0 \frac{f^2}{2}\right] = 0. \qquad (17)$$

Since the expression in the square brackets is a quadratic polynomial in $f$, it is understood that this equation is satisfied if the coefficients of the terms proportional to $f$ and $f^2$ vanish. We thus compose two algebraic equations:

$$\begin{aligned} -a + (-2\mu a + V_0 + 3g\,a) &= 0, \\ \frac{3}{2}a(1 + V_0 + g\,a) &= 0, \end{aligned} \qquad (18)$$

from which we readily obtain

$$a = \frac{-1}{1 + 2(\mu - g)} \qquad (19)$$

and

$$V_0 = -1 + \frac{g}{1 + 2(\mu - g)}. \qquad (20)$$

Since when passing to the linear limit by tending $g \to 0$ the potential depth defined by the last equation becomes equal to $-1$ it is clear that this solution describes the nonlinear transmission resonance $V_{NL1}$ corresponding to the first linear resonance $V_{L1} = -1$.

Analyzing now the obtained formula for the first resonance, we note that the shift from the value of the linear resonance's potential depth is negative for attractive interaction ($g < 0$) and is positive for repulsive interaction ($g > 0$). Interestingly, we note that if the nonlinearity is repulsive and strong enough, the positive shift in the dept of the potential may prevail the first term in Eq. (20) and thus produce a positive value of $V_0 = V_{NL1}$. Hence, in contrast to the linear case, in the nonlinear case reflectionless transmission may occur not only for potential wells but also for potential barriers. This happens when $g > (1+2\mu)/3$. Note finally that it follows from Eq. (20) that for the whole range of considered variation of parameters, i.e., $g \in (-\infty, \mu), \mu > 0$, the maximal range of variation of the potential height for the first transmission resonance is $V_{NL1} \in (-3/2, -1+\mu)$. The dependence of the first resonance position on the nonlinearity parameter $g$ for $\mu = 4$ is shown on Fig. 2.

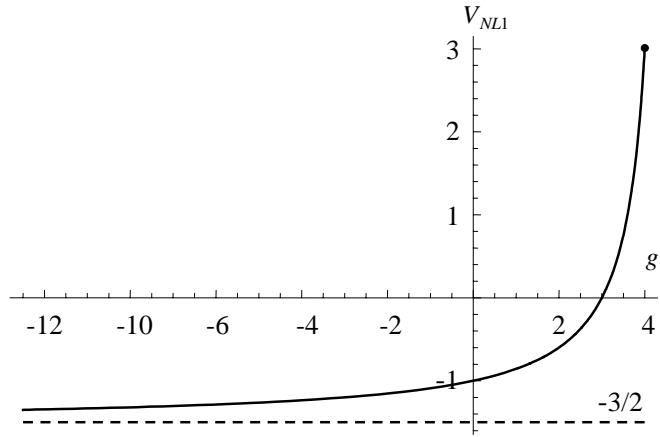

Fig. 2. The position of the first transmission resonance $V_{NL1}$ vs. nonlinearity parameter $g$ for $\mu = 4$. The height of the potential becomes positive starting from $g = (1+2\mu)/3 = 3$.

**Higher order transmission resonances**

To discuss the higher order resonances, we recall that we have previously shown that the position of these resonances in the first-order approximation of the stationary perturbation theory [1] is given as [14]

$$V_{NLn} = V_{Ln} + g F_n / C_n, \qquad (21)$$

where
$$F_n = \int_0^1 z^{ik}(1-z)^{-ik} F(u_{Ln}) u_{Ln}\, dz, \quad F(u) = \frac{|u|^2 - 1}{4z(z-1)} u \qquad (22)$$

and
$$C_n = \int_0^1 z^{ik}(1-z)^{-ik} u_{Lm} u_{Ln}\, dz, \qquad (23)$$

where $u_{Ln}(z)$ is the solution of the linear problem for the $n$ th-order linear resonance given by Eq. (5). It has been numerically proven that these formulas provide a highly accurate description of the nonlinear shift of the position of the reflectionless transmission resonances for all the resonance orders $n$ and all the variation range of the chemical potential $\mu$ if the self-interaction parameter $g$ is relatively small, more precisely, the formulas provide a relative error of the order of $10^{-3}$ if $g/\mu < 0.25$. Notably, the result for the first resonance turns out to be exact, i.e., for $n=1$ Eqs. (21)-(23) produce the exact result (20).

Now, we use the limit solution (13) to estimate the integrals (22) and (23) for higher order resonances $n > 1$. To do this, we construct an approximate solution of Eq. (13) by means of successive iterations. We rewrite this equation in the following form

$$p_0 = \sqrt{\frac{\mu - g}{\mu - V - g\, p_0}} \tag{24}$$

and start from the simplest zero-order initial approximation $p_0 = 1$, i.e., we put $p_0 = 1$ in the right-hand-side of this equation to obtain the first approximation for the limit solution as

$$p_0(z) = \sqrt{\frac{\mu - g}{\mu - g - V(z)}}, \quad V(z) = -V_0 4z(z-1). \tag{25}$$

Numerical testing shows that this is a good enough approximation. Notably it corresponds to the limit solution of a linear problem for the effective chemical potential $\mu - g$. For this function, the integrals (22) and (23) are calculated analytically. The result for $F_n$ is simple:

$$F_n = \frac{-V_{Ln}}{\mu - g - V_{Ln}}, \tag{26}$$

and for $C_n$ we obtain

$$C_n = \frac{1}{2}\sqrt{\frac{\mu - g}{-V_{Ln}}} \ln\left(\frac{V_{Ln} + \sqrt{V_{Ln}(\mu - g)}}{-V_{Ln} + \sqrt{V_{Ln}(\mu - g)}}\right). \tag{27}$$

For fixed wave vector $k = \sqrt{2(\mu - g)}$, the asymptotes of $F_n$ and $C_n$ for large $n \gg 1$ are $F_n \sim 1$ and $C_n \sim 1/\sqrt{-V_{Ln}} \sim 1/n$, hence, we conclude that

$$V_{NLn} - V_{Ln} \sim gn. \tag{28}$$

Thus, the nonlinear shift of the resonance position is approximately a linear function of the resonance order $n$. This conclusion is further supported by numerical simulations (Fig. 3). Recalling now the formula (20) for the first transmission resonance, we finally obtain

$$V_{NLn} \approx V_{Ln} + \frac{g\,n}{1 + 2(\mu - g)}. \tag{29}$$

This formula provides a quick estimate of the nonlinear transmission resonance position and hence may be especially useful for qualitative discussions.

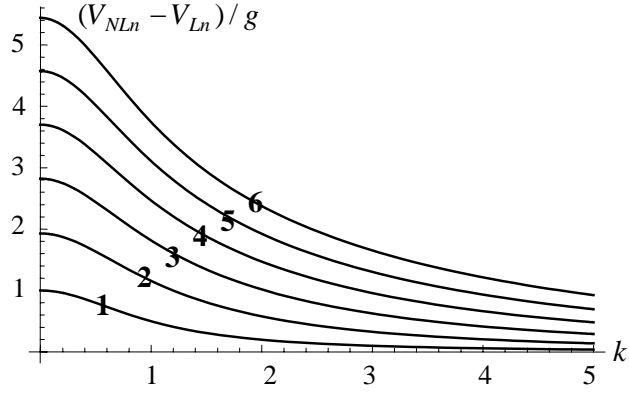

Fig. 3. Rosen-Morse potential: the nonlinear shift of the resonance position $(V_{NLn} - V_{Ln})/g$ as a function of the wave vector $k$. The resonance orders are indicated by bold-face numbers.

It is interesting to compare the obtained result with the one for the rectangular barrier. The latter problem has recently been addressed in several publications (see, e.g. [5]). The result for the transmission resonances presented in these papers is rather cumbersome. Here we present a much simpler discussion due to application of the approach developed in [14]. This approach rests on the reformulation of the reflectionless transmission problem as an eigenvalue problem for the reflecting potential's height and further application of the Rayleigh-Schrödinger time-independent perturbation theory [1]. To proceed in this way, we recall that in the associated *linear* problem for an *effective* chemical potential $\mu_{eff} = \mu - g$ the transmission resonances for the rectangular barrier

$$V = \begin{cases} 0, & x < 0 \text{ and } x > 1 \\ V_0, & 0 \leq x \leq 1 \end{cases} \quad (30)$$

occur for [1]
$$V_{Ln} = (\mu - g) - \frac{\pi^2 n^2}{2}, \quad n = 1, 2, 3, ... \quad (31)$$

(compare with Eq. [4]) and the corresponding wave functions are given as $\psi_{Ln} = e^{-ikx}$ if $x < 0$, $\psi_{Ln} = (-1)^n e^{-ik} e^{-ikx}$ if $x > 1$, and

$$\psi_{Ln} = \cos(\pi n x) - \frac{ik}{\pi n} \sin(\pi n x) \quad \text{if} \quad 0 \leq x \leq 1. \quad (32)$$

[One should put $g = 0$ in Eqs. (30)-(32) if namely the linear problem is considered. However, for the discussion of the nonlinear case that we consider here one should use the formulas with $g \neq 0$.] The functions $\psi_{Ln}$ defined by Eq. (32) are orthogonal in the interval $x \in [0,1]$:

$$\int_0^1 \psi_{Lm}\psi_{Ln}\,dx = C_n\delta_{mn}, \quad C_n = \frac{1}{2} - \frac{k^2}{2\pi^2 n^2}. \tag{33}$$

Then, according to the procedure developed in [14] the nonlinear shift of the resonance position in the limits of the Rayleigh-Schrödinger perturbation theory is given as

$$V_{NLn} - V_{Ln} = \frac{g}{C_n}\int_0^1 (|\psi_{Ln}|^2 - 1)\psi_{Ln}\psi_{Ln}\,dx. \tag{34}$$

.The integral involved in this formula is readily calculated. The final result for the nonlinear resonance position reads

$$V_{NLn} = V_{Ln} + \frac{g}{4}\left(1 - \frac{3k^2}{\pi^2 n^2}\right). \tag{35}$$

This is a fairly good approximation. The numerical simulations show that the derived formula defines the position of the resonances with accuracy of the order of $10^{-4}$ (or less) for all the resonance orders and for all the variation range of the chemical potential $\mu$ and the nonlinearity parameter $g$.

The dependence (35) of the resonance position on the wave vector $k$ of the incoming matter-wave is demonstrated in Fig. 4. It is clearly seen that the situation is radically changed compared with the case of the Rosen-Morse potential (Fig. 3). An immediate observation indicated by Eq. (35) is that for the rectangular barrier (30) the nonlinear shift of the resonance position is approximately constant for higher order resonances $n \gg 1$:

$$V_{NLn} - V_{Ln}(g=0) \approx -\frac{3}{4}g. \tag{36}$$

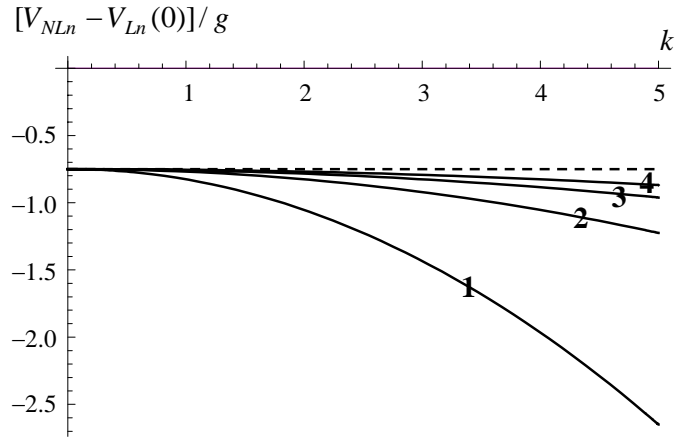

Fig. 4. Rectangular barrier: the nonlinear shift of the resonance position $(V_{NLn} - V_{Ln})/g$ as a function of the wave vector $k$. The bold-face numbers indicate the resonance orders.

Thus, we see that the Rosen-Morse potential suggests an essentially different behavior [Eq. (29)] not indicated by the rectangular barrier [see Eq. (35)]. It is supposed that this is because of the smooth variation of the Rosen-Morse potential's shape. An immediate conjecture following from this observation is that the case of *asymmetric* potentials involving two different scales for the variation of the potential in different space intervals is potent to suggest further effects. We hope to examine this possibility in a future investigation.

**Conclusions**

Thus, we have discussed, within the mean field Gross-Pitaevskii approach, the reflectionless transmission resonances in the above-barrier reflection of Bose-Einstein condensates by the Rosen-Morse squared hyperbolic-secant potential. Applying an exact third order nonlinear differential equation obeyed by the condensate density, we have derived the exact solution of the problem for the first-order resonance. It follows from this solution that in the nonlinear case the total transmission is also possible for positive potential heights, i.e., for potential barriers, not only for the potential wells as it is the case in the linear case. Further, using a limit solution of the equation for the density (written as a root of a polynomial equation of the third degree) and the solution for the first resonance, we have constructed a simple approximation for the shift of the nonlinear resonance potential's depth from the corresponding linear resonance's position for higher order resonances. The constructed approximation shows that the nonlinear shift of the resonance position is an approximately linear function of the resonance order. This behavior essentially differs from the result for the rectangular barrier for which the nonlinear shift is approximately constant. By noting that this radical difference is probably caused by the smooth variation of the Rosen-Morse potential's shape in contrast to the sharp variation in the case of the rectangular barrier, it is conjectured that the case of *asymmetric* potentials involving two different scales for the variation of the potential in different space intervals is potent to suggest further effects. We intend to address this perspective in a forthcoming discussion.

**Acknowledgments**

This research has been conducted in the scope of the International Associated Laboratory IRMAS. The work was supported by the Armenian National Science and Education Fund (ANSEF Grants No. 2009-PS-1692 and No. 2010-MP-2186) and the Russian Foundation for Basic Research (Grant No. 07-02-00080).